\documentclass[aps,
superscriptaddress,
showpacs,
preprintnumbers,
bibnotes,
amsmath,
amssymb,
prd,
floatfix,
]{revtex4-2}
\usepackage[utf8]{inputenc}
\usepackage{array}
\usepackage{subcaption}
\usepackage[normalem]{ulem}
\newcolumntype{L}[1]{>{\raggedright\let\newline\\\arraybackslash\hspace{0pt}}m{#1}}
\newcolumntype{C}[1]{>{\centering\let\newline\\\arraybackslash\hspace{0pt}}m{#1}}
\newcolumntype{R}[1]{>{\raggedleft\let\newline\\\arraybackslash\hspace{0pt}}m{#1}}

\newcommand{\comment}[1]{}

\usepackage{graphicx}
\usepackage{dcolumn}
\usepackage{bm}
\usepackage[usenames,dvipsnames]{color}
\usepackage{amsfonts}
\usepackage{natbib}
\usepackage{xcolor}
\usepackage{hyperref}
\usepackage{dcolumn}

\DeclareMathOperator{\tr}{Tr}
\DeclareMathOperator{\diag}{Diag}

\graphicspath{{figs/}}

\newcommand\regensburg{Fakult\"at f\"ur Physik, Universit\"at Regensburg, Universit\"atsstra{\ss}e 31, 93040 Regensburg, Germany}

\begin{document}
\title{\boldmath Long-distance reconstruction of QED corrections to the hadronic vacuum polarization for the muon g-2}
\author{C.~Lehner}\thanks{Corresponding author}\email{christoph.lehner@ur.de}\affiliation{\regensburg}
\author{J.~Parrino}\affiliation{\regensburg}
\author{A.~V\"olklein}\affiliation{\regensburg}

\noaffiliation

\date{\today}




\begin{abstract}
The long-distance contribution of QED corrections to the hadronic vacuum polarization is particularly challenging to compute in lattice QCD+QED.  Currently, it is one of the limiting factors towards matching the precision of the recent result by the Fermilab E989 experiment for the muon g-2.  In this work, we present a method for obtaining high-precision results for this contribution by reconstructing exclusive finite-volume state contributions.  We find relations between the pion-photon contributions of individual diagrams and demonstrate
the reconstruction method with lattice QCD+QED data at a single lattice spacing of $a^{-1} \approx 1.73$ GeV and $m_\pi \approx 275$ MeV.
\end{abstract}

\maketitle

%


\section{Introduction}
In the recent whitepaper of the Muon g-2 Theory Initiative \cite{Aliberti:2025beg}, lattice QCD+QED calculations have for the first time been used to provide the standard model result for the hadronic vacuum polarization contribution to the muon g-2.  This has been made possible by a concerted effort of the lattice community over the past decade to subdivide the calculation into individual contributions with specific challenges that are then addressed in a targeted approach.  The calculations are typically organized as an expansion around an isospin symmetric QCD calculation, and correctons are added to leading order in the fine-structure constant $\alpha=e^2/(4\pi)$ as well as the mass difference of up- and down-quarks.  The currently available results
\cite{RBC:2018dos,Giusti:2019xct,Borsanyi:2020mff,Boccaletti:2024guq,Djukanovic:2024cmq} unfortunately do not compute the long distance part of the QED corrections from first principles but instead either truncate at some distance or replace the long distance part with a model
such as the phenomenological model of Ref.~\cite{Parrino:2025afq}.

For the isospin symmetric QCD contribution, the challenges of large statistical noise and finite-volume effects at long distances have recently been successfully addressed by a reconstruction of the long-distance part from a dedicated study of exclusive finite-volume states \cite{RBC:2024fic,Djukanovic:2024cmq} completing an effort started many years ago that even included a study of four-pion states \cite{Bruno:2019nzm}.
In the current paper, we extend this idea to QCD+QED calculations by providing a method to reconstruct the long-distance part of the QED corrections to the hadronic vacuum polarization.

In Sec.~\ref{sec:method}, we describe the method and provide a first numerical demonstration in Sec.~\ref{sec:results} for a single lattice spacing of $a^{-1} =1.7312(28)$ GeV, $m_\pi =274.8(2.5)$ MeV, $m_K=530.1(3.1)$ MeV with $m_\pi L = 3.8$ on a $24^3 \times 48 \times 24$ Mobius Domain Wall ensemble with Iwasaki gauge action \cite{RBC:2023pvn,RBC:2024fic}.  We use the QED$_{\rm L}$ \cite{Hayakawa:2008an} but our results directly translate to other QED regulators with a well-defined transfer matrix such as QED$_{\rm r}$ \cite{DiCarlo:2025uyj}.

\section{Methodology}\label{sec:method}
\subsection{QCD + QED correlation functions}
\begin{figure}[tb]
    \centering
    \subcaptionbox*{(c)}{\includegraphics[height=1cm]{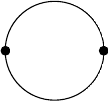}}%
    \hfill
    \subcaptionbox*{(V)}{\includegraphics[height=1cm]{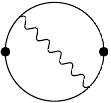}}%
    \hfill
    \subcaptionbox*{(S)}{\includegraphics[height=1cm]{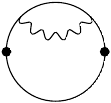}}%
    \hfill
    \subcaptionbox*{(T)}{\includegraphics[height=1cm]{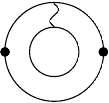}}%
    \hfill
    \subcaptionbox*{(D1)}{\includegraphics[height=1cm]{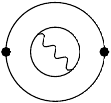}}%
    \hfill
    \subcaptionbox*{(D2)}{\includegraphics[height=1cm]{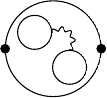}}%
    \hfill
    \\ \vspace{0.5cm}
\subcaptionbox*{(d)}{\includegraphics[height=1cm]{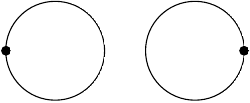}}%
    \hfill
    \subcaptionbox*{(F)}{\includegraphics[height=1cm]{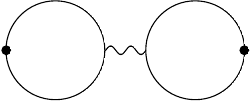}}%
    \hfill
    \subcaptionbox*{(Td)}{\includegraphics[height=1cm]{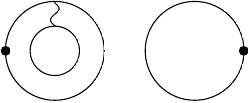}}%
    \hfill
    \subcaptionbox*{(D1d)}{\includegraphics[height=1cm]{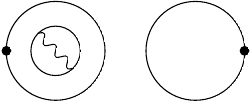}}%
    \hfill
    \subcaptionbox*{(D2d)}{\includegraphics[height=1cm]{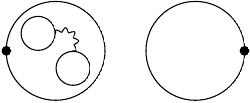}}%
    \hfill
    \subcaptionbox*{(D3)}{\includegraphics[height=1cm]{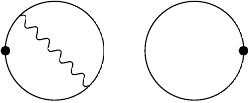}}%
    \hfill
    \\ \vspace{0.5cm}
    \subcaptionbox*{(L)}{\includegraphics[height=1.5cm]{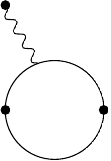}}%
    \hfill
    \subcaptionbox*{(LT)}{\includegraphics[height=1.5cm]{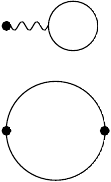}}%
    \hfill
    \subcaptionbox*{(LR)}{\includegraphics[height=1.5cm]{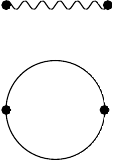}}%
    \hfill
    \subcaptionbox*{(Ld)}{\includegraphics[height=1.5cm]{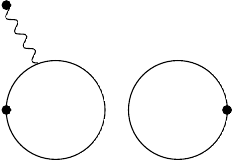}}%
    \hfill
    \subcaptionbox*{(LTd)}{\includegraphics[height=1.5cm]{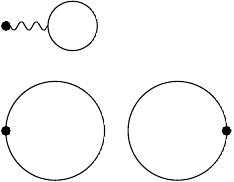}}%
    \hfill
    \caption{QCD + QED diagrams considered in this work.  Diagrams related by  symmetry are identified by their common topology.  The black dots represent the insertion of an external operator.}
    \label{fig:diags}
\end{figure}

Let us consider the spectral representation of the light vector-vector correlator $G(t)$ in Euclidean QCD+QED at second order in the electromagnetic charge $e$,
\begin{align}
    G(t) &\equiv \frac13 \sum_i \langle V_i(t,\vec{p}=0) V^\dagger_i(t=0,\vec{p}=0) \rangle = G^{(0)}(t) + e^2 G^{(2)}(t) 
\end{align} 
with sum over spatial $i$ and
\begin{align}\label{eqn:physicalvector}
     V_j(t,\vec{p}) &= \frac1{\sqrt{L^3}} \sum_{\vec{x}} e^{i \vec{x} \cdot \vec{p}} \left( i \frac{2}{3} \bar u(t,\vec{x}) \gamma_j u(t,\vec{x}) - i\frac{1}{3} \bar d(t,\vec{x}) \gamma_j d(t,\vec{x})\right) \,,
\end{align}
where $u$ and $d$ are the up and down quark fields
and $L^3$ is the spatial volume.
Performing the Wick contractions gives
\begin{align}
\label{eqn:physical}
 G &= \frac59 (c) - \frac19 (d) + e^2 \Bigl(
 -\frac{17}{81}((V) + 2(S))  + \frac{25}{81} (F) + \frac{14}{81} ((T) + (D3))  - \frac{10}{81} (Td)\notag\\&\qquad + \frac{25}{162} (D1)
  - \frac{5}{162}((D1d) + (D2)) + \frac1{162} (D2d)
 \Bigr) 
\end{align}
with diagrams defined in Fig.~\ref{fig:diags}.  Symmetry factors, signs related to fermion loops, and the $-i e$ charge factors are kept explicit throughout this paper.  Suppressing the coordinates, diagram (c) therefore stands for $\tr[ D^{-1} \gamma_i D^{-1} \gamma_i ]$ for a fixed value of spatial $i$.  Since the vector currents transform in the three-dimensional $T_1^{u}$ representation of the octahedral symmetry group, each fixed value of spatial $i$ gives the same result.  We can therefore equivalently consider the diagrams to represent the average over the spatial $i$.

For sufficiently long temporal extents, the correlators can be written in terms of sums over all eigenstates $n$ of the finite-volume Hamiltonian
\begin{align}
G(t) = \sum_n c_n e^{-E_n t} =\sum_n c_n^{(0)} e^{-E_n^{(0)}t} + e^2 \sum_n (c_n^{(2)} - t E_n^{(2)} c_n^{(0)}) e^{-E_n^{(0)}t}
\end{align}
with $c_n = c_n^{(0)} + e^2 c_n^{(2)} = \vert \langle 0 \vert V_i \vert n \rangle \vert^2$ and $E_n = E_n^{(0)} + e^2 E_n^{(2)}$.  Linear terms in $e$ are absent due to the charge-conjugation symmetry of the theory.
The sum over $n$ is over all states of the QCD+QED system, i.e., it includes $\pi^+ \pi^-$ states, a $\rho$ state, as well as states with a photon such as $\pi^0 \gamma$.   States including a photon do not contribute in leading order and, therefore, must have $c_n^{(0)}=0$ in this expression.
The contribution of the $\pi^0 \gamma$ state in the second-order diagrams in $e$ therefore can be written simply as
$e^2 c^{(2)}_{\pi\gamma} e^{-E_{\pi\gamma}^{(0)}t}$.  States that are already present in pure QCD such as the $\rho$ or $\pi\pi$ states receive a correction to both their amplitude and their energy.

The external vector currents need to be renormalized by multiplying appropriate factors of the vector current renormalization factor $Z_V=Z_V^{\rm QCD} Z_V^{\rm QED}$.
If conserved vector currents are used in the lattice calculation, we have $Z_V^{\rm QCD}=1$.  It is useful to write $Z^{\rm QED}_V=1 + e^2 Z_V^{{\rm QED},(2)}$ such that at second order in $e$
the renormalized correlator is
\begin{align}
   G^{\rm ren}(t) = Z_V^2 G(t) &=(Z_V^{\rm QCD})^2\sum_n c_n^{(0)} e^{-E_n^{(0)}t} + e^2 (Z_V^{\rm QCD})^2\sum_n ((c_n^{(2)} + 2Z_V^{{\rm QED},(2)} c_n^{(0)}) - t E_n^{(2)} c_n^{(0)}) e^{-E_n^{(0)}t} \,.
\end{align}
Two additional powers of $Z_V^{\rm QCD}$ for the $e^2$ term need to be included if local currents are used to couple to the photons in the action.  We adopt this choice in this work and therefore additional photon tadpole diagrams are absent in Fig.~\ref{fig:diags}.
The amplitudes $c_n^{(2)}$ for states that are already present in pure QCD are therefore modified by this renormalization procedure, however, the energy corrections $E_n^{(2)}$ and the new state amplitudes such as $c_{\pi\gamma}^{(2)}$ are well-defined even without the inclusion of $Z_V^{\rm QED}$.  Note that additional diagrams to account for shifts of the quark masses need to be included as well to complete the renormalization of the theory but this is not needed for the long-distance reconstruction considered in this work.

In order to reconstruct the long-distance behavior
of $G^{(2)}$, we need to determine $c_n^{(2)}$ and $E_n^{(0)}$
for the new states including a photon as well as
$c_n^{(0)}$, $c_n^{(2)}$, $E_n^{(0)}$, and $E_n^{(2)}$ for the states already present in pure QCD.  This can be achieved by
studying a correlation matrix
\begin{align}
 C_{ij} = \langle O_i O^\dagger_j \rangle \,,
 \end{align}
where one of the operators needs to be $V_i$ and the other operators serve the purpose to project to the states $n$.  In the pure QCD case such a correlation matrix is then used to solve a generalized eigenvalue problem (GEVP)
\begin{align}
 C(t) &= V \diag(e^{-E_1 t},\ldots, e^{-E_n t}) V^\dagger \,, \\
 C(t) \vec{v}_n &= e^{-E_n(t-t_0)} C(t_0) \vec{v}_n
\end{align}
with $V_{in}= \langle 0 \vert O_i \vert n \rangle$.  This is usually studied by considering
\begin{align}
C(t) C(t_0)^{-1} = V \diag(e^{-E_1 (t-t_0)},\ldots, e^{-E_n (t-t_0)}) V^{-1}
\end{align}
and solving for the eigenvalues and eigenvectors of this matrix for sufficiently large $t$ and $t_0$.

In the case of QCD+QED calculations, however,
\begin{align}
 V = V^{(0)} + e V^{(1)} + e^2 V^{(2)} \,,
\end{align}
and a perturbative expansion in $e$ using a wise power-counting scheme for the operators $O_i$ may be more efficient.
To simplify the discussion, we consider only a two-operator system, where the first operator $O_1=V_i$ which serves the purpose to illustrate the pure QCD sector and $O_2=O_{\pi\gamma}$ which should create the lowest-lying pion-gamma state.  Our method extends without loss of generality to the general case with additional multi-pion and pion-photon operators.
In practice, our choice of heavier pion mass for the numerical demonstration in Sec.~\ref{sec:results} makes this two-operator setup already very efficient for the long-distance reconstruction of $G^{(2)}$.  We discuss the construction of the pion-photon operators in Sec.~\ref{sec:pionphoton}.

For sufficiently heavy pion mass the long-distance part of the correlator is dominated by two states $n=\rho$ and $n=\pi\gamma$, such that the correlation matrix becomes
\begin{align}
 C(t) &= \begin{pmatrix} 
 V_{1,\rho} &  V_{1,\pi\gamma} \\
 V_{2,\rho} & V_{2,\pi\gamma} 
 \end{pmatrix}
 \begin{pmatrix} e^{-E_{\rho} t} & 0 \\ 0 & e^{-E_{\pi\gamma} t}
 \end{pmatrix}
\begin{pmatrix} 
 V^*_{1,\rho} &  V^*_{2,\rho} \\
 V^*_{1,\pi\gamma} & V^*_{2,\pi\gamma} 
 \end{pmatrix} \\
 &=\begin{pmatrix}
\vert V_{1,\rho} \vert^2 e^{-E_\rho t} + \vert V_{1,\pi\gamma} \vert^2 e^{-E_{\pi\gamma} t} &  V_{1,\rho} V^*_{2,\rho} e^{-E_\rho t} + V_{1,\pi\gamma} V^*_{2,\pi\gamma} e^{-E_{\pi\gamma} t}\\
V_{2,\rho} V^*_{1,\rho} e^{-E_\rho t} + V_{2,\pi\gamma} V^*_{1,\pi\gamma} e^{-E_{\pi\gamma} t}
& \vert V_{2,\rho} \vert^2 e^{-E_\rho t} + \vert V_{2,\pi\gamma} \vert^2 e^{-E_{\pi\gamma} t}
 \end{pmatrix} \,.
\end{align}
For physical pion mass, two-pion operators and corresponding states need to be added to complete the low-lying spectrum which dominates the long-distance part of $C$.  For the numerical demonstration provided in Sec.~\ref{sec:results},
the lowest two states are the $\rho$ and the $\pi\gamma$ state with smallest relative momentum allowed by the finite volume.

It is convenient to adopt a power-counting in which $O_{\pi\gamma}$ is counted as order $e$, e.g., by multiplying the photon operator with an explicit power of $e$.  We then find
\begin{align}
V = \begin{pmatrix}
V_{1,\rho}^{(0)} + e^2 V_{1,\rho}^{(2)} & e V_{1,\pi\gamma}^{(1)} \\
 e^2 V_{2,\rho}^{(2)} & e V_{2,\pi\gamma}^{(1)}
\end{pmatrix}
\end{align}
and therefore
\begin{align}
 C(t) \label{eqn:corrs}
 &=\begin{pmatrix}
\vert V^{(0)}_{1,\rho} \vert^2 e^{-E^{(0)}_\rho t}  & 0 \\ 0 & 0
 \end{pmatrix} \notag\\&\quad
 + e^2\begin{pmatrix}
 (2 {\rm Re} (V^{(0)}_{1,\rho} V^{(2),*}_{1,\rho}) - t \vert V^{(0)}_{1,\rho} \vert^2 E^{(2)}_\rho) e^{-E^{(0)}_\rho t} + \vert V^{(1)}_{1,\pi\gamma} \vert^2 e^{-E^{(0)}_{\pi\gamma} t} &   V^{(0)}_{1,\rho}  V^{(2),*}_{2,\rho} e^{-E^{(0)}_\rho t} + V^{(1)}_{1,\pi\gamma} V^{(1),*}_{2,\pi\gamma} e^{-E^{(0)}_{\pi\gamma} t}\\
 V^{(2)}_{2,\rho} V^{(0),*}_{1,\rho} e^{-E^{(0)}_\rho t} + V^{(1)}_{2,\pi\gamma} V^{(1),*}_{1,\pi\gamma} e^{-E^{(0)}_{\pi\gamma} t}
&  \vert V^{(1)}_{2,\pi\gamma} \vert^2 e^{-E^{(0)}_{\pi\gamma} t}
 \end{pmatrix} \,.
\end{align}

In the present work, we first determine $E_{\pi\gamma}^{(0)}$
and $\vert V_{2,\pi\gamma}^{(1)} \vert$ from a fit to $C_{22}$
before determining $\vert V^{(1)}_{1,\pi\gamma} \vert$ from a fit to $C_{21}$.  With this knowledge, we subtract the $\pi\gamma$
contribution from $C_{11}$ and fit the subtracted correlator to obtain ${\rm Re} (V^{(0)}_{1,\rho} V^{2,*}_{1,\rho})$ and $E_\rho^{(2)}$.  Combined with $V^{(0)}_{1,\rho}$ and $E^{(0)}_\rho$, which can be obtained from a leading-order GEVP study \cite{RBC:2024fic}, we can then reconstruct both the new $\pi\gamma$ state contribution as well as the QED corrections to the $\rho$ state.

Our method also allows for a separation of the pion-gamma contribution or more general of the new states including a photon such that they can be computed independently from the remainder of the inclusive QED corrections.  This also makes it possible
to subtract the QED$_{\rm L}$ pion-photon contribution and to add it back from an infinite-volume QED calculation.

\subsection{Pion-photon operators}\label{sec:pionphoton}
In this section we discuss the construction of the pion-photon operator for general relative momentum between the pion and photon for the total system at rest.  The construction of a real photon interpolation operator is most straightforward in Coulomb gauge in which unphysical longitudinal contributions do not propagate.  For ease of combination with the usual QCD+QED calculation of the hadronic vacuum polarization correlators, we adopt Feynman gauge in this calculation and benefit from gauge invariance removing the propagation of additional longitudinal states in $G^{(2)}$.
We need to prepare a transverse-polarized photon with momentum $\vec{p}$ combined with a pion state with momentum $-\vec{p}$ and combine the operators such as to create a three-dimensional operator $O_{\pi\gamma,i}$ that transforms in the $T_1^u$ irreducible representation of the octahedral group.

We first define a charged pion operator 
\begin{align}
O_{\pi}(t,\vec{p}) = \frac{i}{\sqrt{L^3}}\sum_{\vec{x}} e^{i\vec{x}\vec{p}} \bar{u}(x) \gamma_5 d(x),
\end{align}
and defer the discussion of the general case to Sec.~\ref{sec:isospin}.  The photon operator is given by
\begin{align}
A_i(t,\vec{p}) = \frac1{\sqrt{L^3}}\sum_{\vec{x}} e^{i\vec{x}\vec{p}} \tilde{A}_i(x)
\end{align}
with photon field $\tilde{A}_i(x)$ of the QED action.
We project to transversal photons by considering
\begin{align}
A^{T}_j(t,\vec{p})&=\Big(\delta_{jm}-\hat{p}_j \hat{p}_m\Big)A_m(t,\vec{p})\,, &
\hat{p}&=\frac{\vec{p}}{|\vec{p}|}\,.
\end{align}

Let $H(\vec{p})$ be the orbit under the chiral octahedral group with 24 elements of a representative momentum vector $\vec{p}$ (e.g., choose the representative with the first nonzero component positive).  Then there is a unique
combination of the operators defined above that transforms in $T_1^u$ of the full octahedral group,
\begin{align}
    \label{eq:general_T1u_op}
O_{\pi\gamma,\vec{p},i}(t)
&=\frac{e}2\sum_{\vec{q} \in H(\vec{p})}\Big(
\big[\hat{q}\times\vec{A}^T(t,\vec{q})\big]_iO_\pi(t,-\vec{q})\;
-\;\big[\hat{q}\times \vec{A}^T(t,-\vec{q})\big]_iO_\pi(t,+\vec{q})
\Big) \\
&=\frac{e}2\sum_{\vec{q} \in H(\vec{p})}\Big(
\big[\hat{q}\times\vec{A}(t,\vec{q})\big]_iO_\pi(t,-\vec{q})\;
-\;\big[\hat{q}\times \vec{A}(t,-\vec{q})\big]_iO_\pi(t,+\vec{q})
\Big)\,.
\end{align}
We note that the projection to the transversal photon is redundant for this operator.
For the smallest allowed momentum we simply find
\begin{align}
 O_{\pi\gamma,100,i}(t) &=\frac{e}2\sum_{j,k} \varepsilon_{ijk} (O_{\pi}(t,\vec{p}=\vert \vec{p} \vert \hat{e}_k) A_j(t,\vec{p}=-\vert \vec{p} \vert \hat{e}_k) - O_{\pi}(t,\vec{p}=-\vert \vec{p} \vert \hat{e}_k) A_j(t,\vec{p}=\vert \vec{p} \vert \hat{e}_k)) 
\end{align}
with $\vert\vec{p}\vert=2\pi/L$ and unit vector $\hat{e}_k$ in the $k$ direction.  We use this operator for $O_{\pi\gamma}$ in Sec.~\ref{sec:results}.

\subsection{Isospin decomposition and relations between diagrams}\label{sec:isospin}
It is instructive to not only consider the physical vector current $V_i$ defined in Eq.~\eqref{eqn:physicalvector}
with quark charges $q_u=2/3$ and $q_d=-1/3$ but instead to consider a general isospin decomposition and charge configurations.  This will allow us to isolate the pion-photon contributions of subsets of diagrams and identify relations between the pion-photon contributions of individual diagrams.  We define
\begin{align}
 O_{\pi(I_3=-1)} &= i\bar u \gamma_5 d \,, & 
 O_{\pi(I_3=0)} &= \frac{i}{\sqrt{2}}\left( \bar u \gamma_5  u - \bar d \gamma_5  d \right) \,, &
 O_{\pi(I_3=1)} &= i\bar d \gamma_5 u \,, \\
 V^j_{I=1,I_3=-1} &= i\bar u \gamma_j d \,, & 
 V^j_{I=1,I_3=0} &= \frac{i}{\sqrt{2}}\left( \bar u \gamma_j u - \bar d \gamma_j d \right) \,,&
 V^j_{I=1,I_3=1} &= i\bar d \gamma_j u \,, \\
 & & V^j_{I=0} &= \frac{i}{\sqrt{2}}\left( \bar u \gamma_j u + \bar d \gamma_j d \right) \,.
\end{align}
We first consider isospin $I=1$ and different charge assignments for the
quark QED charges.  We begin our investigation by keeping the sum of charges constant
\begin{align}
 \langle V_{I=1, I_3=-1} V^\dagger_{I=1, I_3=-1} \rangle_{q_u=1, q_d=1} &= (c) + e^2\left( (D1) - 2 (D2) + 4 (T) - 2 (S) - (V) \right)\,, \\
  \langle V_{I=1, I_3=0} V^\dagger_{I=1, I_3=0} \rangle_{q_u=1, q_d=1} &= (c) + e^2\left((D1) - 2 (D2) + 4 (T) - 2 (S) - (V) \right)\,, \\
  \langle V_{I=1, I_3=-1} V^\dagger_{I=1, I_3=-1} \rangle_{q_u=2, q_d=0} &= (c) + e^2\left(2 (D1) - 2 (D2) + 4 (T) - 4 (S)  \right)\,,\\
  \langle V_{I=1, I_3=0} V^\dagger_{I=1, I_3=0} \rangle_{q_u=2, q_d=0} &= (c) + e^2\left(2 (D1) - 2 (D2) + 4 (T) - 4 (S) + 2 (F) - 2 (V) \right)
\end{align}
such that the correlators including the pion-photon operator
remain identical, i.e.,
\begin{align}
\langle O_{\pi(I_3=-1),\gamma} V^\dagger_{I=1, I_3=-1} \rangle_{q_u=1, q_d=1} &= e^2 \left(-2i (LT) + 2i (L)\right)\,, \\
\langle O_{\pi(I_3=0),\gamma} V^\dagger_{I=1, I_3=0} \rangle_{q_u=1, q_d=1} &= e^2 \left(-2i (LT) + 2i (L)\right)\,, \\
\langle O_{\pi(I_3=-1),\gamma} V^\dagger_{I=1, I_3=-1} \rangle_{q_u=2, q_d=0} &=e^2 \left( -2i (LT) + 2i (L)\right)\,, \\
\langle O_{\pi(I_3=0),\gamma} V^\dagger_{I=1, I_3=0} \rangle_{q_u=2, q_d=0} &=e^2 \left( -2i (LT) + 2i (L) \right) \,, \\
\langle O_{\pi(I_3=-1),\gamma} O_{\pi(I_3=-1),\gamma}^\dagger \rangle &= e^2 (LR) \,.
\end{align}

From this, we  infer that the pion-photon contributions, which only depend on $|V_{1,\pi\gamma}^{\left(1\right)}|^{2}$
determined by the unchanged pion-photon operator correlators,
obey the relations
\begin{align}
    (F) &= (V) \,, &  2(S) - (D1) = (V) \,. \label{eqn:pion-photon-relations}
\end{align}
Based on these relations, we could construct correlators in which the pion-photon
contribution is absent as has been done for the hadronic light-by-light contribution's pion pole \cite{Blum:2023vlm}.  In the present work, however, it is our intention to reconstruct this contribution instead by studying the enlarged operator basis.

If we let the charges sum to zero instead, the off-diagonal correlators vanish,
\begin{align}
    \langle O_{\pi(I_3=-1),\gamma} V^\dagger_{I=1, I_3=-1} \rangle_{q_u=1, q_d=-1} &= 0 \,,\\
\langle O_{\pi(I_3=0),\gamma} V^\dagger_{I=1, I_3=0} \rangle_{q_u=1, q_d=-1} &= 0
\end{align}
and
\begin{align}
\langle V_{I=1, I_3=-1} V^\dagger_{I=1, I_3=-1} \rangle_{q_u=1, q_d=-1} &= (c) + e^2\left((D1)  - 2 (S) + (V) \right)\,, \\
  \langle V_{I=1, I_3=0} V^\dagger_{I=1, I_3=0} \rangle_{q_u=1, q_d=-1} &= (c) + e^2\left((D1) - 2 (S) + 2 (F) - (V) \right) \,.
\end{align}
The pion-photon contributions need to cancel in these
linear combinations of diagrams and using Eqs.~\eqref{eqn:pion-photon-relations}, we see that this is indeed the case.

Next, we consider the $I=0$ case for which an isospin symmetric charge assignment yields
\begin{align}\label{eqn:isozero}
 \langle V_{I=0} V^\dagger_{I=0} \rangle_{q_u=1, q_d=1} &= (c)-2(d) + e^2\Bigl( 4(T) - 8(Td) -2(D1d) + 4(D3) + 4(D2d) - 2(D2) + (D1) \notag\\
 &\qquad + 2(F) - (V) - 2(S)\Bigr) \,, \\
    \langle O_{\pi(I_3=0),\gamma} V^\dagger_{I=0} \rangle_{q_u=1, q_d=1} &= 0
\end{align}
and the pion-photon contribution in the linear combination of diagrams in Eq.~\eqref{eqn:isozero}
needs to vanish.



Combined with Eq.~\eqref{eqn:pion-photon-relations} the pion-photon relations can be written as
\begin{align}\label{eqn:pion-photon-relations-all}
    (F)&=(V) \,, & 2(S) - (D1) & = (V) \,, &
      2(T) - 4(Td) -(D1d) + 2(D3) + 2(D2d) - (D2) &= 0 \,.
\end{align}

If we collect the pion photon contribution from diagrams (V), (S), (D1), and (F) in the total hadronic vacuum polarization contribution given in Eq.~\eqref{eqn:physical}, we find
\begin{align}
 -e^2 \frac{9}{162} ((V) + 2(S)) \,.
\end{align}
We observe that the pion-photon contribution is strongly suppressed
due to a cancellation between the (F), (D1), (S) and (V) diagrams.
To obtain the total contribution of the pion-photon states created by the physical $V_i^\dagger$,
we need
\begin{align}
 \langle O_{\pi(I_3=0),\gamma} V^\dagger \rangle &= e^2 \left( -\frac{\sqrt{2}}{6}i ((LT) + (Ld)) + \frac{\sqrt{2}}{3}i (L) \right) \,.
\end{align}


\section{Results}\label{sec:results}
In this section, we provide a numerical demonstration of our approach.  We use data at a single lattice spacing of $a^{-1} =1.7312(28)$ GeV, $m_\pi =274.8(2.5)$ MeV, $m_K=530.1(3.1)$ MeV with $m_\pi L = 3.8$ on a $24^3 \times 48 \times 24$ Mobius Domain Wall ensemble with Iwasaki gauge action.  This corresponds to ``ensemble 4'' of Refs.~\cite{RBC:2023pvn,RBC:2024fic}.
For this pion mass, the lowest two states are the $\rho$ and the $\pi\gamma$ state with lowest relative momentum.  We proceed as outlined in Sec.~\ref{sec:method}.

\begin{figure}[tb]
    \centering
    \includegraphics[page=1,scale=0.8]{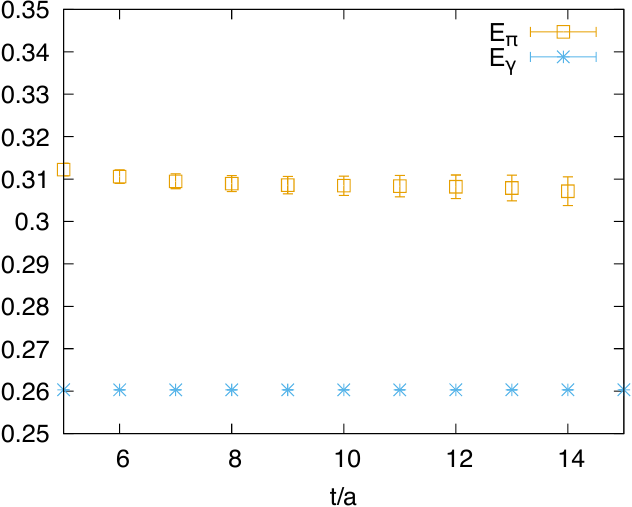}
    \hfill
    \includegraphics[page=2,scale=0.8]{figs/C1-LR.pdf}
    \hfill
    \caption{Analysis of diagram (LR).  The left plot shows the effective masses separately obtained for the photon and pion factors of the diagram.  The right plot shows the fitted normalization of the pion-photon state with fit range [t,24].}
    \label{fig:diagLR}
\end{figure}

\begin{figure}[tb]
    \centering
    \includegraphics[page=1,scale=0.8]{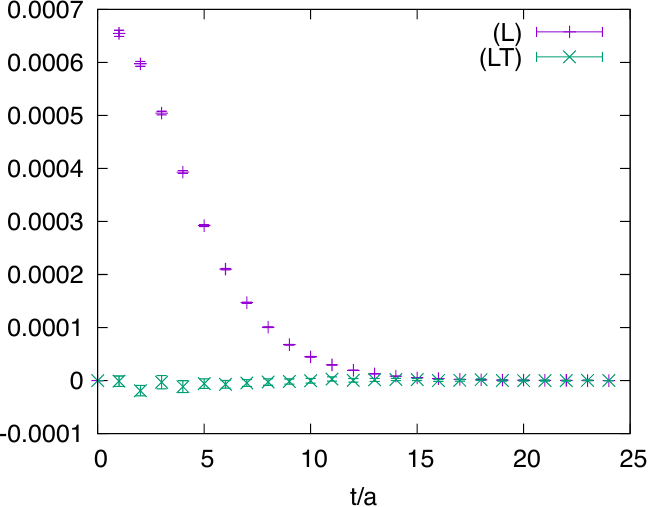}%
    \hfill
    \includegraphics[page=2,scale=0.8]{figs/C1-L.pdf}%
    \hfill
    \caption{Analysis of diagrams (L) and (LT).  The diagram (LT) is numerically negligible compared to (L) as shown in the left plot.  On the right side, we show the fitted overlap factors to the $\rho$ and $\pi\gamma$ states with fit range [t,24].}
    \label{fig:diagNL}
\end{figure}

In Fig.~\ref{fig:diagLR}, we show results for $E_{\pi\gamma}^{(0)}=E_{\pi} + E_\gamma$
and $\vert V_{2,\pi\gamma}^{(1)} \vert$ from a fit to $C_{22}$.
With these values fixed, we determine $\vert V^{(1)}_{1,\pi\gamma}\vert$ from a fit to $C_{21}$ as shown in the right panel of Fig.~\ref{fig:diagNL}.  The left panel of this figure shows the contribution of the individual diagrams to $C_{21}$ we note that
diagram (LT) is numerically negligible compared to (L).

\begin{figure}[tb]
    \centering
\includegraphics[page=1,scale=0.8]{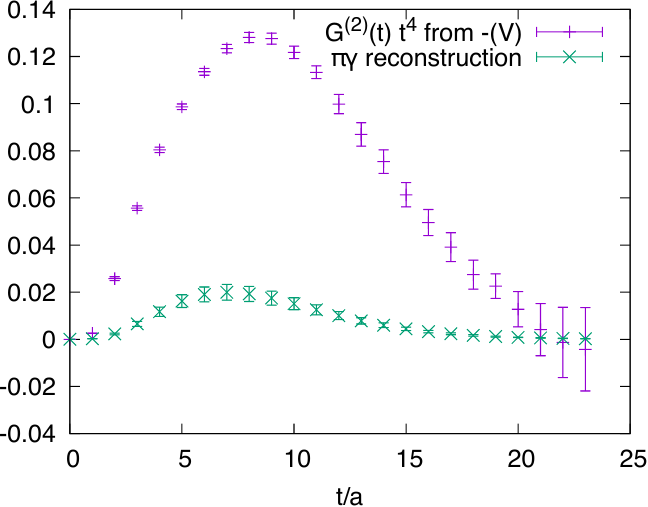}%
    \hfill
    \includegraphics[page=2,scale=0.8]{figs/all.pdf}%
    \hfill \\
    \includegraphics[page=3,scale=0.8]{figs/all.pdf}%
    \hfill
    \includegraphics[page=4,scale=0.8]{figs/all.pdf}%
    \hfill
    \caption{Study of $t^4$ integrand and the $\pi\gamma$ reconstruction for diagrams (V), (S), and (F).}
    \label{fig:diagVSF}
\end{figure}

Using the relations of Eq.~\ref{eqn:pion-photon-relations-all},
we reconstruct the pion-gamma contribution to individual diagrams (V), (S), and (F) assuming dominance over diagram (D1) for now. In Fig.~\ref{fig:diagVSF}, we compare this reconstruction to the individual diagrams for a $t^4$ integrand which behaves similarly as the time-momentum representation kernel \cite{Bernecker:2011gh} for the muon g-2.  We note that the combination $(V) + 2(S)$, which also appears for the hadronic vacuum polarization \ref{eqn:physical}, is well approximated by the pion-photon state.  We note, however, that without performing the vector current renormalization as described in Sec.~\ref{sec:method}, this comparison is incomplete.

\begin{figure}[tb]
    \centering
    \includegraphics[page=5,scale=0.8]{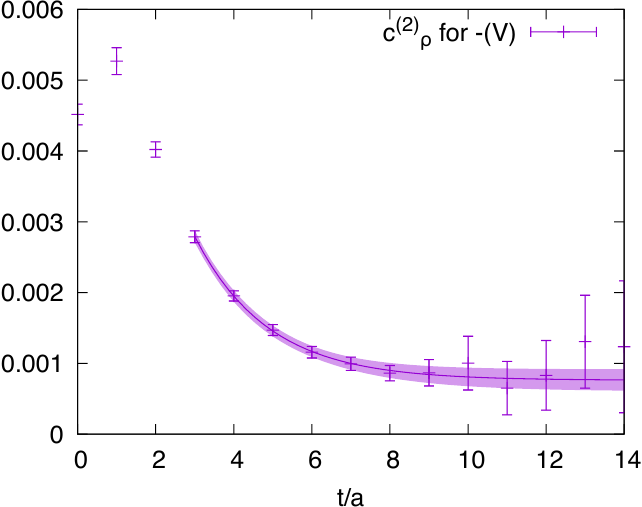}%
    \hfill
    \includegraphics[page=6,scale=0.8]{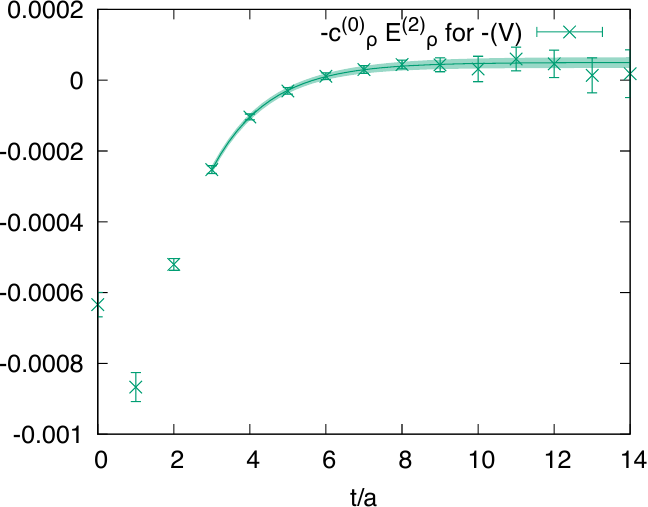}%
    \hfill \\
    \includegraphics[page=9,scale=0.8]{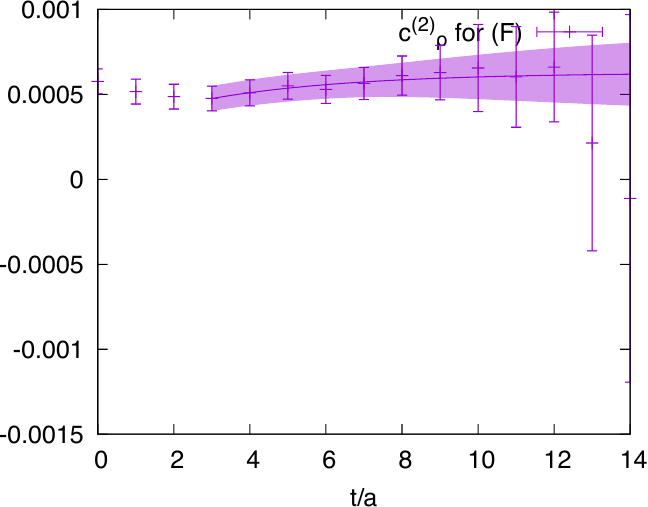}%
    \hfill
    \includegraphics[page=10,scale=0.8]{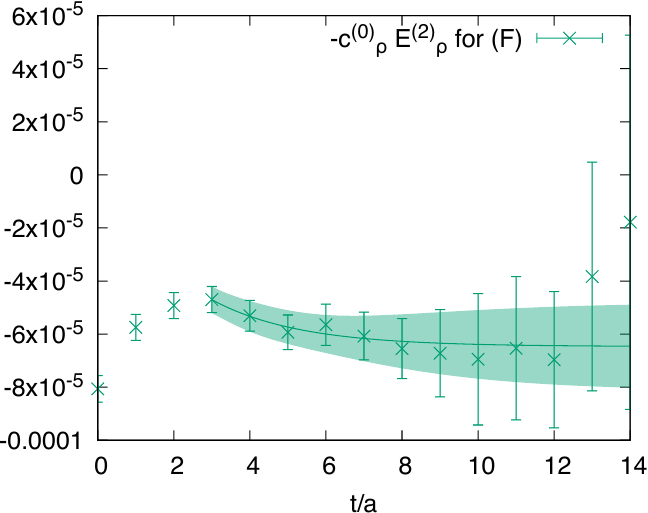}%
    \hfill \\
    \includegraphics[page=7,scale=0.8]{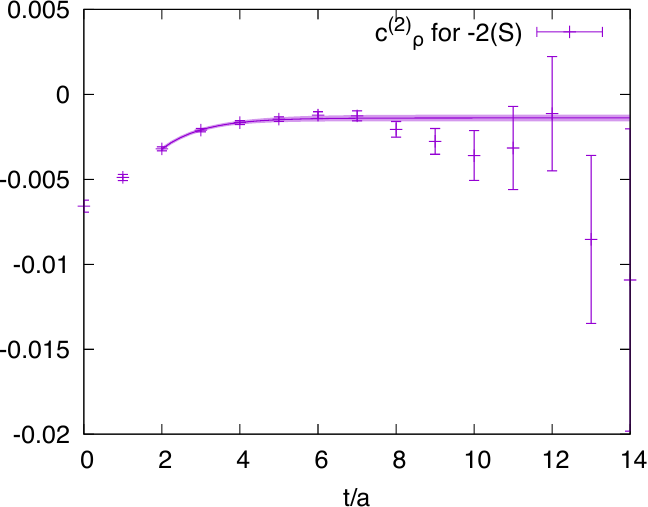}%
    \hfill
    \includegraphics[page=8,scale=0.8]{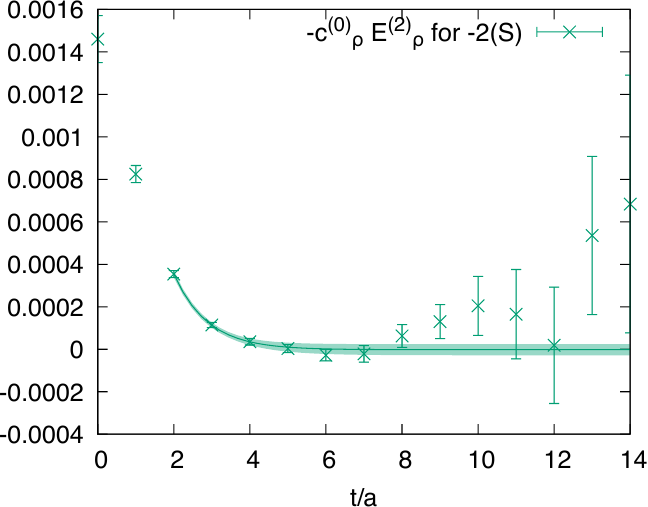}%
    \hfill
    \caption{Fits for the QED corrections to $\rho$ parameters.}
    \label{fig:fitQEDRHO}
\end{figure}

\begin{figure}[tb]
    \centering
    \subcaptionbox*{(V)}{\includegraphics[page=11,scale=0.8]{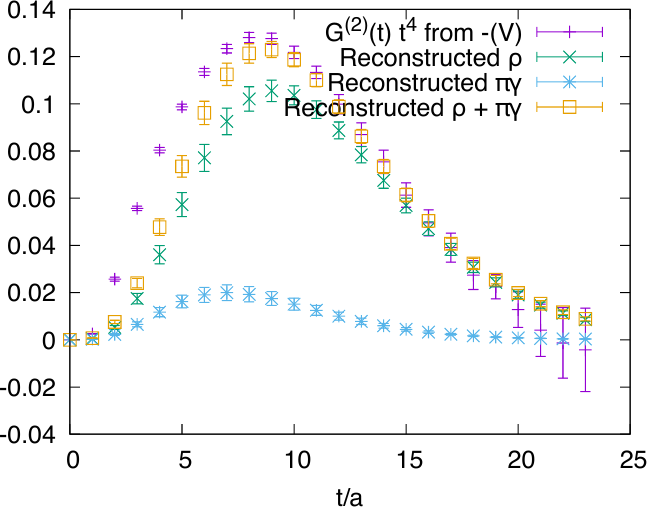}}%
    \hfill
    \subcaptionbox*{(L)} {\includegraphics[page=14,scale=0.8]{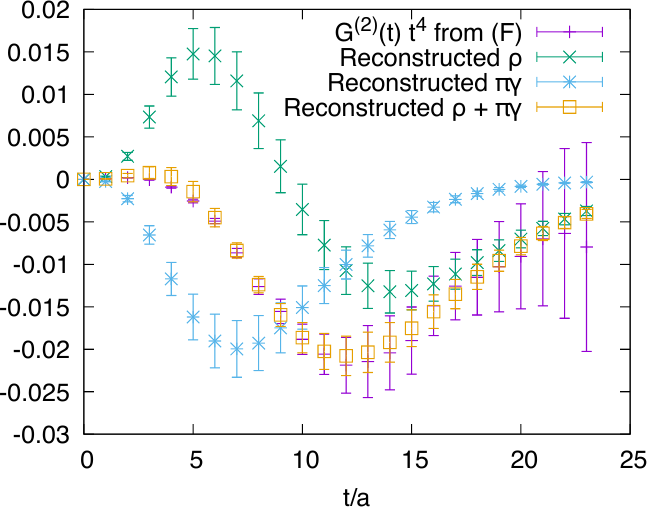}}%
    \hfill \\
    \subcaptionbox*{(V)}{\includegraphics[page=12,scale=0.8]{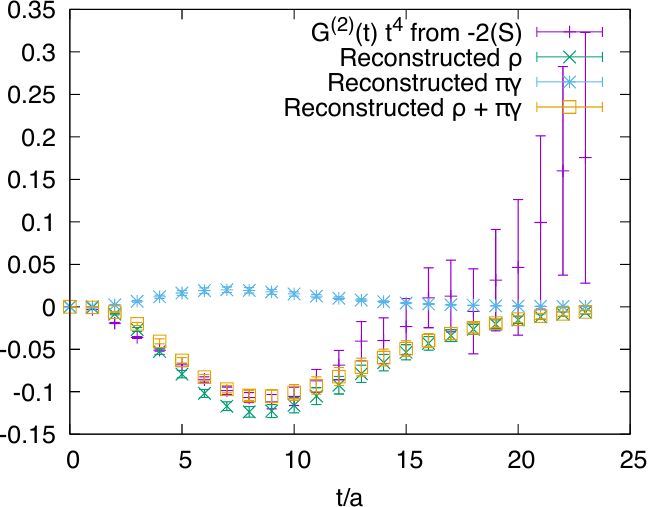}}%
    \hfill
    \subcaptionbox*{(L)} {\includegraphics[page=13,scale=0.8]{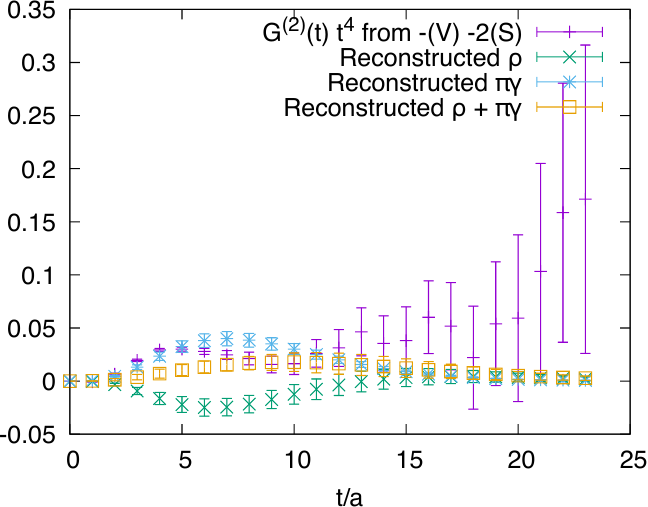}}%
    \hfill
    \caption{Full reconstruction for diagrams (V), (S), and (F).}
    \label{fig:reconst}
\end{figure}

With the data obtained so far, we subtract the $\pi\gamma$
contribution from $C_{11}$ and fit the subtracted correlator to obtain $c_\rho^{(2)} = 2 {\rm Re}(V^{(0)}_{1,\rho} V^{(2),*}_{1,\rho})$ and $c_\rho^{(0)} E_\rho^{(2)} = \vert V^{(0)}_{1,\rho} \vert^2 E_\rho^{(2)}$. 
In this fit, we also use the information for $V^{(0)}_{1,\rho}$ and $E^{(0)}_\rho$ from the leading-order GEVP study \cite{RBC:2024fic} for this ensemble.  We show the resulting fits to these $\rho$ corrections in Fig.~\ref{fig:fitQEDRHO}.
We perform the fits including excited state contributions accounting amongst others for the higher-lying two-pion states.
The fits perform well even for small time-slices and allow us to reconstruct both the $\rho$ and $\pi\gamma$ contributions to the individual diagrams.  We show the results for this reconstruction in Fig.~\ref{fig:reconst}.

\begin{figure}[tb]
    \centering
    \subcaptionbox*{(V)}{\includegraphics[page=15,scale=0.7]{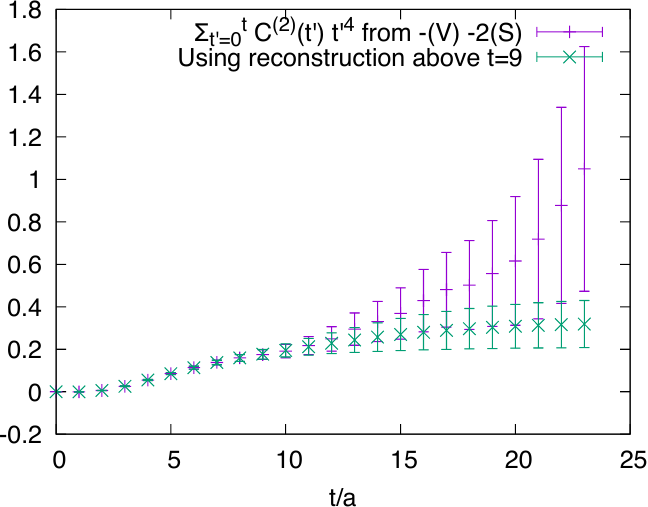}}%
    \hfill
    \subcaptionbox*{(L)} {\includegraphics[page=16,scale=0.7]{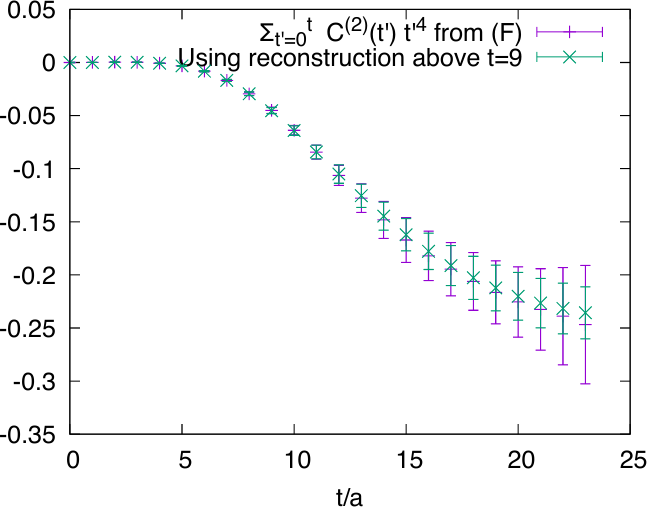}}%
    \hfill
    \caption{Partial sum of the $G^{(2)}(t) t^4$ integrand that closely resembles the time-momentum kernel for the hadronic vacuum polarization contribution to the muon g-2.}
    \label{fig:partialsum}
\end{figure}

In Fig.~\ref{fig:partialsum}, we show the partial sum of the $G^{(2)}(t)t^4$ integrand which closely resembles the desired time-momentum integrand for the QED corrections to the hadronic vacuum polarization contribution to the muon g-2 \cite{Bernecker:2011gh}. We observe a reduction of statistical noise by more than a factor of five for the noisy diagram (S).

 %
 %
 %
 %
 %






%
%

\section{Conclusion}
In this paper, we have described a method for a high-precision first-principles calculation of the long-distance tail of the QED corrections to the hadronic vacuum polarization contribution for the muon g-2.  We found relations between the pion-photon contributions of individual diagrams which allows for the dominant diagrams (V), (S), and (F) to be analyzed individually after accounting for the contribution of (D1).  
 We also observed a substantial suppression factor of the pion-photon contribution to the hadronic vacuum polarization because of a cancelation of these individual contributions.
 Our numerical results at approximately twice the physical pion mass demonstrate the potential for noise reduction of our approach.  For the noisy diagram (S), the reduction of uncertainty was observed to be greater than a factor of five.
In future work, we will extend our studies to different photon regulators including a photon in infinite volume.  We will also include our existing data set of multi-pion operators and physical pion mass ensembles.  Finally, we will include the sub-leading diagrams that we have neglected so far.

\section{Acknowledgments}
We thank our colleagues of the RBC and UKQCD collaborations for many valuable discussions and joint efforts over the years.
The authors gratefully acknowledge the Gauss Centre for Supercomputing
e.V. (www.gauss-centre.eu) for funding this project by providing
computing time on the GCS Supercomputer JUWELS at Jülich
Supercomputing Centre (JSC).
We
gratefully acknowledge disk and tape storage provided by USQCD and by
the University of Regensburg with support from the DFG.
The lattice data analyzed in this project was generated using GPT
\cite{GPT} and Grid \cite{GRID,Boyle:2016lbp,Yamaguchi:2022feu}.

\bibliography{references}

\begin{thebibliography}{18}%
\makeatletter
\providecommand \@ifxundefined [1]{%
 \@ifx{#1\undefined}
}%
\providecommand \@ifnum [1]{%
 \ifnum #1\expandafter \@firstoftwo
 \else \expandafter \@secondoftwo
 \fi
}%
\providecommand \@ifx [1]{%
 \ifx #1\expandafter \@firstoftwo
 \else \expandafter \@secondoftwo
 \fi
}%
\providecommand \natexlab [1]{#1}%
\providecommand \enquote  [1]{``#1''}%
\providecommand \bibnamefont  [1]{#1}%
\providecommand \bibfnamefont [1]{#1}%
\providecommand \citenamefont [1]{#1}%
\providecommand \href@noop [0]{\@secondoftwo}%
\providecommand \href [0]{\begingroup \@sanitize@url \@href}%
\providecommand \@href[1]{\@@startlink{#1}\@@href}%
\providecommand \@@href[1]{\endgroup#1\@@endlink}%
\providecommand \@sanitize@url [0]{\catcode `\\12\catcode `\$12\catcode
  `\&12\catcode `\#12\catcode `\^12\catcode `\_12\catcode `\%12\relax}%
\providecommand \@@startlink[1]{}%
\providecommand \@@endlink[0]{}%
\providecommand \url  [0]{\begingroup\@sanitize@url \@url }%
\providecommand \@url [1]{\endgroup\@href {#1}{\urlprefix }}%
\providecommand \urlprefix  [0]{URL }%
\providecommand \Eprint [0]{\href }%
\providecommand \doibase [0]{https://doi.org/}%
\providecommand \selectlanguage [0]{\@gobble}%
\providecommand \bibinfo  [0]{\@secondoftwo}%
\providecommand \bibfield  [0]{\@secondoftwo}%
\providecommand \translation [1]{[#1]}%
\providecommand \BibitemOpen [0]{}%
\providecommand \bibitemStop [0]{}%
\providecommand \bibitemNoStop [0]{.\EOS\space}%
\providecommand \EOS [0]{\spacefactor3000\relax}%
\providecommand \BibitemShut  [1]{\csname bibitem#1\endcsname}%
\let\auto@bib@innerbib\@empty
\bibitem [{\citenamefont {Aliberti}\ \emph {et~al.}(2025)\citenamefont
  {Aliberti} \emph {et~al.}}]{Aliberti:2025beg}%
  \BibitemOpen
  \bibfield  {author} {\bibinfo {author} {\bibfnamefont {R.}~\bibnamefont
  {Aliberti}} \emph {et~al.},\ }\bibfield  {title} {\bibinfo {title} {{The
  anomalous magnetic moment of the muon in the Standard Model: an update}},\
  }\href@noop {} {\  (\bibinfo {year} {2025})},\ \Eprint
  {https://arxiv.org/abs/2505.21476} {arXiv:2505.21476 [hep-ph]} \BibitemShut
  {NoStop}%
\bibitem [{\citenamefont {Blum}\ \emph {et~al.}(2018)\citenamefont {Blum},
  \citenamefont {Boyle}, \citenamefont {G\"ulpers}, \citenamefont {Izubuchi},
  \citenamefont {Jin}, \citenamefont {Jung}, \citenamefont {J\"uttner},
  \citenamefont {Lehner}, \citenamefont {Portelli},\ and\ \citenamefont
  {Tsang}}]{RBC:2018dos}%
  \BibitemOpen
  \bibfield  {author} {\bibinfo {author} {\bibfnamefont {T.}~\bibnamefont
  {Blum}}, \bibinfo {author} {\bibfnamefont {P.~A.}\ \bibnamefont {Boyle}},
  \bibinfo {author} {\bibfnamefont {V.}~\bibnamefont {G\"ulpers}}, \bibinfo
  {author} {\bibfnamefont {T.}~\bibnamefont {Izubuchi}}, \bibinfo {author}
  {\bibfnamefont {L.}~\bibnamefont {Jin}}, \bibinfo {author} {\bibfnamefont
  {C.}~\bibnamefont {Jung}}, \bibinfo {author} {\bibfnamefont {A.}~\bibnamefont
  {J\"uttner}}, \bibinfo {author} {\bibfnamefont {C.}~\bibnamefont {Lehner}},
  \bibinfo {author} {\bibfnamefont {A.}~\bibnamefont {Portelli}},\ and\
  \bibinfo {author} {\bibfnamefont {J.~T.}\ \bibnamefont {Tsang}} (\bibinfo
  {collaboration} {RBC, UKQCD}),\ }\bibfield  {title} {\bibinfo {title}
  {{Calculation of the hadronic vacuum polarization contribution to the muon
  anomalous magnetic moment}},\ }\href
  {https://doi.org/10.1103/PhysRevLett.121.022003} {\bibfield  {journal}
  {\bibinfo  {journal} {Phys. Rev. Lett.}\ }\textbf {\bibinfo {volume} {121}},\
  \bibinfo {pages} {022003} (\bibinfo {year} {2018})},\ \Eprint
  {https://arxiv.org/abs/1801.07224} {arXiv:1801.07224 [hep-lat]} \BibitemShut
  {NoStop}%
\bibitem [{\citenamefont {Giusti}\ \emph {et~al.}(2019)\citenamefont {Giusti},
  \citenamefont {Lubicz}, \citenamefont {Martinelli}, \citenamefont
  {Sanfilippo},\ and\ \citenamefont {Simula}}]{Giusti:2019xct}%
  \BibitemOpen
  \bibfield  {author} {\bibinfo {author} {\bibfnamefont {D.}~\bibnamefont
  {Giusti}}, \bibinfo {author} {\bibfnamefont {V.}~\bibnamefont {Lubicz}},
  \bibinfo {author} {\bibfnamefont {G.}~\bibnamefont {Martinelli}}, \bibinfo
  {author} {\bibfnamefont {F.}~\bibnamefont {Sanfilippo}},\ and\ \bibinfo
  {author} {\bibfnamefont {S.}~\bibnamefont {Simula}} (\bibinfo {collaboration}
  {ETM}),\ }\bibfield  {title} {\bibinfo {title} {{Electromagnetic and strong
  isospin-breaking corrections to the muon $g - 2$ from Lattice QCD+QED}},\
  }\href {https://doi.org/10.1103/PhysRevD.99.114502} {\bibfield  {journal}
  {\bibinfo  {journal} {Phys. Rev. D}\ }\textbf {\bibinfo {volume} {99}},\
  \bibinfo {pages} {114502} (\bibinfo {year} {2019})},\ \Eprint
  {https://arxiv.org/abs/1901.10462} {arXiv:1901.10462 [hep-lat]} \BibitemShut
  {NoStop}%
\bibitem [{\citenamefont {Bors{\'a}nyi}\ \emph {et~al.}(2021)\citenamefont
  {Bors{\'a}nyi} \emph {et~al.}}]{Borsanyi:2020mff}%
  \BibitemOpen
  \bibfield  {author} {\bibinfo {author} {\bibfnamefont {S.}~\bibnamefont
  {Bors{\'a}nyi}} \emph {et~al.},\ }\bibfield  {title} {\bibinfo {title}
  {{Leading hadronic contribution to the muon magnetic moment from lattice
  QCD}},\ }\href {https://doi.org/10.1038/s41586-021-03418-1} {\bibfield
  {journal} {\bibinfo  {journal} {Nature}\ }\textbf {\bibinfo {volume} {593}},\
  \bibinfo {pages} {51} (\bibinfo {year} {2021})},\ \Eprint
  {https://arxiv.org/abs/2002.12347} {arXiv:2002.12347 [hep-lat]} \BibitemShut
  {NoStop}%
\bibitem [{\citenamefont {Boccaletti}\ \emph {et~al.}(2024)\citenamefont
  {Boccaletti} \emph {et~al.}}]{Boccaletti:2024guq}%
  \BibitemOpen
  \bibfield  {author} {\bibinfo {author} {\bibfnamefont {A.}~\bibnamefont
  {Boccaletti}} \emph {et~al.},\ }\bibfield  {title} {\bibinfo {title} {{High
  precision calculation of the hadronic vacuum polarisation contribution to the
  muon anomaly}},\ }\href@noop {} {\  (\bibinfo {year} {2024})},\ \Eprint
  {https://arxiv.org/abs/2407.10913} {arXiv:2407.10913 [hep-lat]} \BibitemShut
  {NoStop}%
\bibitem [{\citenamefont {Djukanovic}\ \emph {et~al.}(2025)\citenamefont
  {Djukanovic}, \citenamefont {von Hippel}, \citenamefont {Kuberski},
  \citenamefont {Meyer}, \citenamefont {Miller}, \citenamefont {Ottnad},
  \citenamefont {Parrino}, \citenamefont {Risch},\ and\ \citenamefont
  {Wittig}}]{Djukanovic:2024cmq}%
  \BibitemOpen
  \bibfield  {author} {\bibinfo {author} {\bibfnamefont {D.}~\bibnamefont
  {Djukanovic}}, \bibinfo {author} {\bibfnamefont {G.}~\bibnamefont {von
  Hippel}}, \bibinfo {author} {\bibfnamefont {S.}~\bibnamefont {Kuberski}},
  \bibinfo {author} {\bibfnamefont {H.~B.}\ \bibnamefont {Meyer}}, \bibinfo
  {author} {\bibfnamefont {N.}~\bibnamefont {Miller}}, \bibinfo {author}
  {\bibfnamefont {K.}~\bibnamefont {Ottnad}}, \bibinfo {author} {\bibfnamefont
  {J.}~\bibnamefont {Parrino}}, \bibinfo {author} {\bibfnamefont
  {A.}~\bibnamefont {Risch}},\ and\ \bibinfo {author} {\bibfnamefont
  {H.}~\bibnamefont {Wittig}},\ }\bibfield  {title} {\bibinfo {title} {{The
  hadronic vacuum polarization contribution to the muon g \ensuremath{-} 2 at
  long distances}},\ }\href {https://doi.org/10.1007/JHEP04(2025)098}
  {\bibfield  {journal} {\bibinfo  {journal} {JHEP}\ }\textbf {\bibinfo
  {volume} {04}},\ \bibinfo {pages} {098}},\ \Eprint
  {https://arxiv.org/abs/2411.07969} {arXiv:2411.07969 [hep-lat]} \BibitemShut
  {NoStop}%
\bibitem [{\citenamefont {Parrino}\ \emph {et~al.}(2025)\citenamefont
  {Parrino}, \citenamefont {Biloshytskyi}, \citenamefont {Chao}, \citenamefont
  {Meyer},\ and\ \citenamefont {Pascalutsa}}]{Parrino:2025afq}%
  \BibitemOpen
  \bibfield  {author} {\bibinfo {author} {\bibfnamefont {J.}~\bibnamefont
  {Parrino}}, \bibinfo {author} {\bibfnamefont {V.}~\bibnamefont
  {Biloshytskyi}}, \bibinfo {author} {\bibfnamefont {E.-H.}\ \bibnamefont
  {Chao}}, \bibinfo {author} {\bibfnamefont {H.~B.}\ \bibnamefont {Meyer}},\
  and\ \bibinfo {author} {\bibfnamefont {V.}~\bibnamefont {Pascalutsa}},\
  }\bibfield  {title} {\bibinfo {title} {{Computing the UV-finite
  electromagnetic corrections to the hadronic vacuum polarization in the muon
  (g {\ensuremath{-}} 2) from lattice QCD}},\ }\href
  {https://doi.org/10.1007/JHEP07(2025)201} {\bibfield  {journal} {\bibinfo
  {journal} {JHEP}\ }\textbf {\bibinfo {volume} {07}},\ \bibinfo {pages}
  {201}},\ \Eprint {https://arxiv.org/abs/2501.03192} {arXiv:2501.03192
  [hep-lat]} \BibitemShut {NoStop}%
\bibitem [{\citenamefont {Blum}\ \emph
  {et~al.}(2025{\natexlab{a}})\citenamefont {Blum} \emph
  {et~al.}}]{RBC:2024fic}%
  \BibitemOpen
  \bibfield  {author} {\bibinfo {author} {\bibfnamefont {T.}~\bibnamefont
  {Blum}} \emph {et~al.} (\bibinfo {collaboration} {RBC, UKQCD}),\ }\bibfield
  {title} {\bibinfo {title} {{The long-distance window of the hadronic vacuum
  polarization for the muon g-2}},\ }\href
  {https://doi.org/10.1103/PhysRevLett.134.201901} {\bibfield  {journal}
  {\bibinfo  {journal} {Phys. Rev. Lett.}\ }\textbf {\bibinfo {volume} {134}},\
  \bibinfo {pages} {201901} (\bibinfo {year} {2025}{\natexlab{a}})},\ \Eprint
  {https://arxiv.org/abs/2410.20590} {arXiv:2410.20590 [hep-lat]} \BibitemShut
  {NoStop}%
\bibitem [{\citenamefont {Bruno}\ \emph {et~al.}(2019)\citenamefont {Bruno},
  \citenamefont {Izubuchi}, \citenamefont {Lehner},\ and\ \citenamefont
  {Meyer}}]{Bruno:2019nzm}%
  \BibitemOpen
  \bibfield  {author} {\bibinfo {author} {\bibfnamefont {M.}~\bibnamefont
  {Bruno}}, \bibinfo {author} {\bibfnamefont {T.}~\bibnamefont {Izubuchi}},
  \bibinfo {author} {\bibfnamefont {C.}~\bibnamefont {Lehner}},\ and\ \bibinfo
  {author} {\bibfnamefont {A.~S.}\ \bibnamefont {Meyer}},\ }\bibfield  {title}
  {\bibinfo {title} {{Exclusive Channel Study of the Muon HVP}},\ }\href
  {https://doi.org/10.22323/1.363.0239} {\bibfield  {journal} {\bibinfo
  {journal} {PoS}\ }\textbf {\bibinfo {volume} {LATTICE2019}},\ \bibinfo
  {pages} {239} (\bibinfo {year} {2019})},\ \Eprint
  {https://arxiv.org/abs/1910.11745} {arXiv:1910.11745 [hep-lat]} \BibitemShut
  {NoStop}%
\bibitem [{\citenamefont {Blum}\ \emph {et~al.}(2023)\citenamefont {Blum} \emph
  {et~al.}}]{RBC:2023pvn}%
  \BibitemOpen
  \bibfield  {author} {\bibinfo {author} {\bibfnamefont {T.}~\bibnamefont
  {Blum}} \emph {et~al.} (\bibinfo {collaboration} {RBC, UKQCD}),\ }\bibfield
  {title} {\bibinfo {title} {{Update of Euclidean windows of the hadronic
  vacuum polarization}},\ }\href {https://doi.org/10.1103/PhysRevD.108.054507}
  {\bibfield  {journal} {\bibinfo  {journal} {Phys. Rev. D}\ }\textbf {\bibinfo
  {volume} {108}},\ \bibinfo {pages} {054507} (\bibinfo {year} {2023})},\
  \Eprint {https://arxiv.org/abs/2301.08696} {arXiv:2301.08696 [hep-lat]}
  \BibitemShut {NoStop}%
\bibitem [{\citenamefont {Hayakawa}\ and\ \citenamefont
  {Uno}(2008)}]{Hayakawa:2008an}%
  \BibitemOpen
  \bibfield  {author} {\bibinfo {author} {\bibfnamefont {M.}~\bibnamefont
  {Hayakawa}}\ and\ \bibinfo {author} {\bibfnamefont {S.}~\bibnamefont {Uno}},\
  }\bibfield  {title} {\bibinfo {title} {{QED in finite volume and finite size
  scaling effect on electromagnetic properties of hadrons}},\ }\href
  {https://doi.org/10.1143/PTP.120.413} {\bibfield  {journal} {\bibinfo
  {journal} {Prog. Theor. Phys.}\ }\textbf {\bibinfo {volume} {120}},\ \bibinfo
  {pages} {413} (\bibinfo {year} {2008})},\ \Eprint
  {https://arxiv.org/abs/0804.2044} {arXiv:0804.2044 [hep-ph]} \BibitemShut
  {NoStop}%
\bibitem [{\citenamefont {Di~Carlo}\ \emph {et~al.}(2025)\citenamefont
  {Di~Carlo}, \citenamefont {Hansen}, \citenamefont {Hermansson-Truedsson},\
  and\ \citenamefont {Portelli}}]{DiCarlo:2025uyj}%
  \BibitemOpen
  \bibfield  {author} {\bibinfo {author} {\bibfnamefont {M.}~\bibnamefont
  {Di~Carlo}}, \bibinfo {author} {\bibfnamefont {M.~T.}\ \bibnamefont
  {Hansen}}, \bibinfo {author} {\bibfnamefont {N.}~\bibnamefont
  {Hermansson-Truedsson}},\ and\ \bibinfo {author} {\bibfnamefont
  {A.}~\bibnamefont {Portelli}},\ }\bibfield  {title} {\bibinfo {title}
  {{QED$_\text{r}$: a finite-volume QED action with redistributed spatial
  zero-momentum modes}},\ }\href@noop {} {\  (\bibinfo {year} {2025})},\
  \Eprint {https://arxiv.org/abs/2501.07936} {arXiv:2501.07936 [hep-lat]}
  \BibitemShut {NoStop}%
\bibitem [{\citenamefont {Blum}\ \emph
  {et~al.}(2025{\natexlab{b}})\citenamefont {Blum}, \citenamefont {Christ},
  \citenamefont {Hayakawa}, \citenamefont {Izubuchi}, \citenamefont {Jin},
  \citenamefont {Jung}, \citenamefont {Lehner},\ and\ \citenamefont
  {Tu}}]{Blum:2023vlm}%
  \BibitemOpen
  \bibfield  {author} {\bibinfo {author} {\bibfnamefont {T.}~\bibnamefont
  {Blum}}, \bibinfo {author} {\bibfnamefont {N.}~\bibnamefont {Christ}},
  \bibinfo {author} {\bibfnamefont {M.}~\bibnamefont {Hayakawa}}, \bibinfo
  {author} {\bibfnamefont {T.}~\bibnamefont {Izubuchi}}, \bibinfo {author}
  {\bibfnamefont {L.}~\bibnamefont {Jin}}, \bibinfo {author} {\bibfnamefont
  {C.}~\bibnamefont {Jung}}, \bibinfo {author} {\bibfnamefont {C.}~\bibnamefont
  {Lehner}},\ and\ \bibinfo {author} {\bibfnamefont {C.}~\bibnamefont {Tu}}
  (\bibinfo {collaboration} {RBC, UKQCD}),\ }\bibfield  {title} {\bibinfo
  {title} {{Hadronic light-by-light contribution to the muon anomaly from
  lattice QCD with infinite volume QED at physical pion mass}},\ }\href
  {https://doi.org/10.1103/PhysRevD.111.014501} {\bibfield  {journal} {\bibinfo
   {journal} {Phys. Rev. D}\ }\textbf {\bibinfo {volume} {111}},\ \bibinfo
  {pages} {014501} (\bibinfo {year} {2025}{\natexlab{b}})},\ \Eprint
  {https://arxiv.org/abs/2304.04423} {arXiv:2304.04423 [hep-lat]} \BibitemShut
  {NoStop}%
\bibitem [{\citenamefont {Bernecker}\ and\ \citenamefont
  {Meyer}(2011)}]{Bernecker:2011gh}%
  \BibitemOpen
  \bibfield  {author} {\bibinfo {author} {\bibfnamefont {D.}~\bibnamefont
  {Bernecker}}\ and\ \bibinfo {author} {\bibfnamefont {H.~B.}\ \bibnamefont
  {Meyer}},\ }\bibfield  {title} {\bibinfo {title} {{Vector Correlators in
  Lattice QCD: Methods and applications}},\ }\href
  {https://doi.org/10.1140/epja/i2011-11148-6} {\bibfield  {journal} {\bibinfo
  {journal} {Eur. Phys. J. A}\ }\textbf {\bibinfo {volume} {47}},\ \bibinfo
  {pages} {148} (\bibinfo {year} {2011})},\ \Eprint
  {https://arxiv.org/abs/1107.4388} {arXiv:1107.4388 [hep-lat]} \BibitemShut
  {NoStop}%
\bibitem [{\citenamefont {{C. Lehner et al.}}()}]{GPT}%
  \BibitemOpen
  \bibfield  {author} {\bibinfo {author} {\bibnamefont {{C. Lehner et al.}}},\
  }\href {https://github.com/lehner/gpt} {\bibinfo {title} {{Grid Python
  Toolkit (GPT)}}}\BibitemShut {NoStop}%
\bibitem [{\citenamefont {{P.A. Boyle et al.}}()}]{GRID}%
  \BibitemOpen
  \bibfield  {author} {\bibinfo {author} {\bibnamefont {{P.A. Boyle et al.}}},\
  }\href {https://github.com/paboyle/Grid} {\bibinfo {title}
  {{Grid}}}\BibitemShut {NoStop}%
\bibitem [{\citenamefont {Boyle}\ \emph {et~al.}(2016)\citenamefont {Boyle},
  \citenamefont {Cossu}, \citenamefont {Yamaguchi},\ and\ \citenamefont
  {Portelli}}]{Boyle:2016lbp}%
  \BibitemOpen
  \bibfield  {author} {\bibinfo {author} {\bibfnamefont {P.~A.}\ \bibnamefont
  {Boyle}}, \bibinfo {author} {\bibfnamefont {G.}~\bibnamefont {Cossu}},
  \bibinfo {author} {\bibfnamefont {A.}~\bibnamefont {Yamaguchi}},\ and\
  \bibinfo {author} {\bibfnamefont {A.}~\bibnamefont {Portelli}},\ }\bibfield
  {title} {\bibinfo {title} {{Grid: A next generation data parallel C++ QCD
  library}},\ }\href {https://doi.org/10.22323/1.251.0023} {\bibfield
  {journal} {\bibinfo  {journal} {PoS}\ }\textbf {\bibinfo {volume}
  {LATTICE2015}},\ \bibinfo {pages} {023} (\bibinfo {year} {2016})}\BibitemShut
  {NoStop}%
\bibitem [{\citenamefont {Yamaguchi}\ \emph {et~al.}(2022)\citenamefont
  {Yamaguchi}, \citenamefont {Boyle}, \citenamefont {Cossu}, \citenamefont
  {Filaci}, \citenamefont {Lehner},\ and\ \citenamefont
  {Portelli}}]{Yamaguchi:2022feu}%
  \BibitemOpen
  \bibfield  {author} {\bibinfo {author} {\bibfnamefont {A.}~\bibnamefont
  {Yamaguchi}}, \bibinfo {author} {\bibfnamefont {P.}~\bibnamefont {Boyle}},
  \bibinfo {author} {\bibfnamefont {G.}~\bibnamefont {Cossu}}, \bibinfo
  {author} {\bibfnamefont {G.}~\bibnamefont {Filaci}}, \bibinfo {author}
  {\bibfnamefont {C.}~\bibnamefont {Lehner}},\ and\ \bibinfo {author}
  {\bibfnamefont {A.}~\bibnamefont {Portelli}},\ }\bibfield  {title} {\bibinfo
  {title} {{Grid: OneCode and FourAPIs}},\ }\href
  {https://doi.org/10.22323/1.396.0035} {\bibfield  {journal} {\bibinfo
  {journal} {PoS}\ }\textbf {\bibinfo {volume} {LATTICE2021}},\ \bibinfo
  {pages} {035} (\bibinfo {year} {2022})},\ \Eprint
  {https://arxiv.org/abs/2203.06777} {arXiv:2203.06777 [hep-lat]} \BibitemShut
  {NoStop}%
\end{thebibliography}%

\end{document}